\documentclass[preprin,showpacs,preprintnumbers,eqsecnum,amsmath,amssymb,nofootinbib]{revtex4}
\usepackage{graphics,epsfig}
\usepackage{graphicx}
\usepackage{dcolumn}
\usepackage{bm}

\begin{document}
\baselineskip=0.8 cm
\title{ Thermodynamical study on universal horizons in higher $D$-dimensional spacetime and aether waves}

\author{Chikun Ding }
\email{Chikun_Ding@huhst.edu.cn; dingchikun@163.com}
  \affiliation{ Department of Physics, Hunan University of Humanities, Science and Technology, Loudi, Hunan
417000, P. R. China\\
Key Laboratory of Low Dimensional
Quantum Structures and Quantum Control of Ministry of Education,
and Synergetic Innovation Center for Quantum Effects and Applications,
Hunan Normal University, Changsha, Hunan 410081, P. R. China}

\author{ Anzhong Wang}
 \email{Anzhong_Wang@baylor.edu}
 \affiliation{GCAP-CASPER, Physics Department, Baylor University, Waco, Texas 76798-7316, USA\\
Institute for Advanced Physics $\&$ Mathematics, Zhejiang University of Technology, Hangzhou, 310032, China}

\vspace*{0.2cm}
\begin{abstract}
\baselineskip=0.6 cm
\begin{center}
{\bf Abstract}
\end{center}

We investigate thermodynamic behaviors of the $D$-dimensional  gravity coupled to a dynamical unit timelike vector, the aether, present two kinds of exact charged solutions and study the linearized wave spectrum of this theory.  There is an universal horizon behind the Killing horizon in these aether black holes. It is interesting that there exist $D$-dimensional Schwarzschild, Reissner-Nordstr\"{o}m and charged Schwarzschild black holes but now with an universal horizon inside. We find that in the uncharged case for both kinds, or charged case but the charge $\bar Q\ll \bar r_0/2$ for the first kind, one can construct a Smarr formula and the (slightly modified) first law of black hole mechanics at the universal horizons for the aether black holes. An entropy can be associated with the universal horizon and a temperature can be defined there.
For the second kind aether black hole and for the first kind one in the extremal higher dimensions, the work term involving charge $V_{UH}\delta Q$ disappeared. For aether wave, our results show that the spin-1 and spin-2 modes are the same as those in 4-dimensional spacetime, and only the spin-0 one is different and dependant on the dimension number $n$.

\end{abstract}

 \maketitle

\vspace*{0.2cm}
\section{Introduction}

Lorentz invariance(LI) is one of the fundamental principles of GR and the standard model(SM) of particles and fields. However, LI should not be an exact symmetry at all energies  \cite{mattingly}, particularly when one considering quantum gravity effect, it should not be applicable. Though both GR and SM based on LI and the background of spacetime, they handle their entities in profoundly different manners. GR is a classical field theory in curved spacetime that neglects all quantum properties of particles; SM is a quantum field theory in flat spacetime that neglects all gravitational effects of particles. For collisions of particles of $10^{30}$ eV energy (energy higher than Planck scale), the gravitational interactions predicted by GR are very strong and gravity should not be negligible\cite{camelia}. So in this very high energy scale,  one have to consider merging SM with GR in a single unified theory, known as ``quantum gravity", which remains a challenging task. Lorentz symmetry is a continuous spacetime symmetry and cannot exist in a discrete spacetime. Therefore quantization of spacetime at energies beyond the Planck energy, Lorentz symmetry is invalid and one should reconsider giving up LI. There are some phenomena of LV. On the SM side, there is an {\it a priori }unknown physics at high-energy scales that could lead to a spontaneous breaking of LI by giving an expectation value to certain non-SM fields that carry Lorentz indices\cite{bolokhov}. LI also leads to divergences in quantum field theory which can be cured with a short distance of cutoff that breaks it \cite{jacobson}. On the GR side, astrophysical observations suggest that the high-energy cosmic rays above the Greisen-Zatsepin-Kuzmin cutoff are a result of LV\cite{carroll}.

 Thus, the study of LV is a valuable tool to probe the foundations of modern physics. These studies include LV in the neutrino sector \cite{dai}, the standard-model extension \cite{colladay}, LV in the non-gravity sector \cite{coleman}, and LV effect on the formation of atmospheric showers \cite{rubtsov}.
Einstein-aether theory can be considered as an effective description of Lorentz symmetry breaking in the gravity sector and has been extensively used in order to obtain quantitative constraints on Lorentz-violating gravity\cite{jacobson2}.

In Einstein-aether theory, the background tensor fields $u^a$ break the Lorentz symmetry only down to a rotation subgroup by the existence of a preferred time direction at every point of spacetime. The introduction of the aether vector preserves general covariance, however, allows for some novel effects, e.g., matter fields can travel faster than the speed of light \cite{jacobson3}, dubbed superluminal particle. Due the existence of the superluminal gravitational modes, so the corresponding light-cones can be completely flat, and the causality is more like that of Newtonian theory\cite{greenwald}.
In the static, spherically symmetric spacetime, there still exists a causal boundary---universal horizon that can trap excitations traveling at arbitrarily high velocities. The universal horizon locates at $(u\cdot\chi)=0$, where the Killing vector $\chi^a$ becomes tangent to $\Sigma_U$ hypersurface which orthogonal to $u^a$. Any signal  must propagate to the future in aether time $U$, which is towards decreasing  $r$ and generated by $u^a$, at the universal horizon.
 Recently, two exact charged black hole solutions and their Smarr formula on universal horizons in 4- and 3-dimensional spacetime were found by Ding {\it et al} \cite{ding,ding2}. Constraints on Einstein-aether theory were studied by Oost {\it et al} \cite{oost} and, gravitational wave studied by Gong {\it et al} \cite{gong2018} after GW170817. Other studies on universal horizons can be found in \cite{tian}.

In the higher than 4-dimensional spacetime, there is a work on the aether black hole solutions \cite{lin2017} done by Kai Lin and Hei-Hung Ho. However, its thermodynamic study is absent. In the present paper, we study the thermodynamic properties of higher $D$-dimensional solutions and the linearized wave spectrum in this spacetime.

There are several motivations for study on higher than four dimensional black holes. If the purpose of GR is to describe the geometry of spacetime, then it should not be restricted only to 4-dimensions \cite{pereniguez}. Endowing GR with a tunable parameter $D$ should lead to valuable insight into the nature of the theory \cite{emparan}. String theory contains gravity and requires more than four dimensions and it achieves to a successful  microstate description of black hole entropy. Brane World model suggested that our four dimension universe is actually a membrane embedded in a five dimensional spacetime. The AdS/CFT correspondence relates the dynamics of a $D$-dimensional black hole with those of a quantum field theory in $D-1$ dimensions without gravity. In the large $D$ limit, recent study shows that the theory of GR becomes extremely simplified: black holes behave as non-interacting particles \cite{emparan2013}.

The rest of the paper is organized as follows. In Sec. II we provide the background for the Einstein-Maxwell-aether theory studied in this paper. In Sec. III we review the Smarr formula of a black hole, and then show the procedure of how to construct a Smarr formula for spherically symmetric solutions. In Sec. IV, we first construct  two new classes of exact charged solutions, and then use them as examples to study the Smarr formula and  first law. In Sec. V, we present our main conclusions and some discussions. In Appendix A and B, we study the linearized Einstein-aether theory and give some detailed derivations for formulas in maintext in $D$-dimensional spacetime.

\section{Einstein-Maxwell-aether theory}

The general action for the Einstein-Maxwell-aether theory can be constructed by assuming that: (1) it is general covariant; and (2)  it  is a functional of only the spacetime metric $g_{ab}$, an unit timelike vector $u^a$ and Maxwell field $\mathcal{A}^a$, and involves  no more than two derivatives of them. So that the resulting field equations are second-order differential equations of  $g_{ab}$,   $u^a$ and $\mathcal{A}^a$. To simplify the problem, the coupling between aether field and Maxwell one is ignored. Then,  the  Einstein-Maxwell-aether theory to be studied   in this paper is
 described by the  action,
\begin{eqnarray}
\mathcal{S}=
\int d^Dx\sqrt{-g}\Big[\frac{1}{16\pi G_{\ae}}(\mathcal{R}-2\Lambda+\mathcal{L}_{\ae}+\mathcal{L}_M)\Big]\,, \label{action}
\end{eqnarray}
where $D\geq4$ is the number of dimensions of the given spacetime. To a spherically symmetrical spacetime, one can split the dimensions $D$ into $D=n+2$ with coordinates $(t,r,\theta_1,\theta_2,\cdots,\theta_n)$, whose transverse space is a $n$ dimensional sphere space.

In terms of the tensor $Z^{ab}_{~~cd}$ defined as \cite{eling,garfinkle},
\begin{eqnarray}
Z^{ab}_{~~cd}=c_1g^{ab}g_{cd}+c_2\delta^a_{~c}\delta^b_{~d}
+c_3\delta^a_{~d}\delta^b_{~c}-c_4u^au^bg_{cd}\,,
\end{eqnarray}
 the aether Lagrangian $\mathcal{L}_{\ae}$ is given by
\begin{eqnarray}
-\mathcal{L}_{\ae}&=&Z^{ab}_{~~cd}(\nabla_au^c)(\nabla_bu^d)-\lambda(u^2+1)\nonumber\\
&=&c_1(\nabla_au_d)(\nabla^au^d)+c_2(\nabla_au^a)^2+c_3(\nabla_au^b)(\nabla_bu^a)
-c_4a^2-\lambda(u^2+1),
\end{eqnarray}
where $c_i (i = 1, 2, 3, 4)$ are coupling constants of the theory and $a_d=u^a\nabla_au_d$.
The aether Lagrangian is therefore the sum of all possible terms for the aether field $u^a$ up to mass dimension two, and the constraint term $\lambda(u^2 + 1)$ with the Lagrange multiplier $\lambda$ implementing the normalization condition $u^2=-1$.
The source-free Maxwell Lagrangian $\mathcal{L}_M$ is given by
\begin{equation}
\mathcal{L}_M=-\frac{1}{4}\mathcal{F}_{ab}\mathcal{F}^{ab},
~\mathcal{F}_{ab}=\nabla_a\mathcal{A}_b-\nabla_b\mathcal{A}_a,
\end{equation}
where $\mathcal{A}_a$ is the electromagnetic potential four-vector.

Here,  we impose the following
constraints (see Appendix for more detail),
\begin{equation}
\label{CDs}
0\leq c_{14}<\frac{n}{n-1},\quad 1+c_2+\frac{n-1}{n}c_{123}>0,\quad 0\leq c_{13}<1,
\end{equation}
where $c_{14}\equiv c_1+c_4$, and so on\footnote{There are more constrains, see Eqs. (2.18) in \cite{lin2018}}.

The equations of motion, obtained by varying  the action (\ref{action}) with respect to $g_{ab}$, $u^a$, $\mathcal{A}^a$ and $\lambda$ are
\begin{eqnarray}\label{motion}
\mathcal{G}_{ab}+\Lambda g_{ab}=\mathcal{T}^{\ae}_{ab}+8\pi G_{\ae}\mathcal{T}^M_{ab},\quad {\AE}_a=0,\quad \nabla ^a\mathcal{F}_{ab}=0, \quad u^2=-1,
\end{eqnarray}
respectively, where the aether and Maxwell energy-momentum stress tensors $\mathcal{T}^{\ae}_{ab}$ (for its derivation, see Appendix B) and $\mathcal{T}^M_{ab}$ are given by
\begin{eqnarray}
\label{EMTs}
&&\mathcal{T}^{\ae}_{ab}=\lambda u_au_b+c_4a_aa_b-\frac{1}{2}g_{ab}Y^c_{~~d}\nabla_cu^d+\nabla_cX^c_{~~ab}
+c_1[(\nabla_au_c)(\nabla_bu^c)-(\nabla^cu_a)(\nabla_cu_b)],\nonumber \\
&&\mathcal{T}_{ab}^M
 =\frac{1}{16\pi G_{\ae}}\Big[-\frac{1}{4}g_{ab}\mathcal{F}_{mn}\mathcal{F}^{mn}
 +\mathcal{F}_{am}\mathcal{F}_{b}^{~m}\Big],
 \end{eqnarray}
with
\begin{eqnarray}
{\AE}_a=\nabla_bY^b_{~~a}+\lambda u_a+c_4(\nabla_au^b)a_b,\quad
Y^a_{~~b}=Z^{ac}_{~~~bd}\nabla_cu^d, \quad
 X^c_{~~ab}=Y^c_{~~(a}u_{b)}-u_{(a}Y^{~~c}_{b)}+u^cY_{(ab)} .
\end{eqnarray}
The acceleration vector $a^a$ appearing in the expression for the aether energy-momentum stress tensor is defined as the parallel transport of the aether field along itself, $a^a
\equiv \nabla_uu^a,$ where $\nabla_X\equiv X^b\nabla_b$.

Following \cite{berglund}, in spherically symmetry spacetime, the symmetry enforces $u^a$ hypersurface-orthogonal and becoming normal to one or more constant-radius hyperfurface that lies inside the Killing horizon. So one can let $\Sigma_U$ denote a surface orthogonal to the aether vector $u^a$, then $U$ is the aether time generated by $u^a$ that specifies each hypersurface in a foliation. As one moves in toward the origin, each $\Sigma_U$ hypersurface bends down to the infinite past, and asymptoting to a ($D-1$)-dimensional spacelike hypersurface on which $(u\cdot\chi)=0$, which implies that the Killing vector $\chi^a$ becomes tangent to $\Sigma_U$. This hypersurface is the universal horizon. Therefore, we are going to reduce these equations to a spherical symmetry case.

  We first define a set of basis vectors at every point in the spacetime,
so that we can project out various components of the equations of motion.
Let us first  take the aether field $u^a$ to be the basis vector. Then, pick up $n$ spacelike unit vectors, denoted, respectively, by $m_{(i)}^a$ $(i=1,2,\cdots,n)$, all of which are normalized to unity,  mutually orthogonal, and lie on the tangent plane of the $n$-spheres $\mathcal{B}$ that foliate the hypersurface $\Sigma_U$. Finally, let us pick up $s^a$, a spacelike unit vector that is orthogonal to $u^a$, $m_{(i)}^a$ and  points ``outwards'' along a $\Sigma_U$ hypersurface, so we have the frame tetrad, $(u^a, s^a, m_{(i)}^a)$,  with the metric
\begin{eqnarray}
 ds^2 = -u_au_b + s_as_b + \hat{g}_{ab},
\end{eqnarray}
 where $\hat{g}_{ab}$ is $n$-spheres $\mathcal{B}$ of $D$-dimensional spherically spacetime.

 By spherical symmetry, any physical vector $A^a$ has at most two non-vanishing components along, respectively,  $u^a$ and $s^a$, i.e., $A^a=A_1u^a+A_2s^a$.  In particular,
 the acceleration $a^a$  has only one component along $s^a$, namely,  $a^a=(a\cdot s)s^a$.
Similarly, any rank-two tensor $F_{ab}$ may have components along the directions of the bi-vectors $u_au_b,~u_{(a}s_{b)},~u_{[a}s_{b]},~s_as_b,~\hat{g}_{ab}$. The most basic expansions among these basis are the derivatives for  $u^a$ and $s^a$
\begin{eqnarray}
&& \nabla_au_b=-u_aa_b+K_{ab},\;K_{ab}=K_0s_as_b+\frac{\hat{K}}{n}\hat{g}_{ab},\nonumber\\
&& \nabla_as_b=K_0s_au_b+K^{(s)}_{ab},\;K^{(s)}_{ab}=-(a\cdot s)u_au_b+\frac{\hat{k}}{n}\hat{g}_{ab},
\end{eqnarray}
where $K_{ab}^{(s)}$ is the extrinsic curvature of the hypersurfaces $\Sigma_s$, and $\hat{k}$ and $\hat{K}$ are the traces of the extrinsic curvatures of a $n$-sphere $\mathcal{B}$
\begin{eqnarray}\label{usTa}
\hat{k} \equiv \frac{1}{n} g^{ab} {\cal{L}}_s\hat g_{ab},\;\;\;
\hat{K} \equiv \frac{1}{n} g^{ab} {\cal{L}}_u\hat g_{ab},
\end{eqnarray}
with $K=\nabla\cdot u \; (\equiv K_0+\hat{K}) $ being  the trace of the extrinsic curvature of the  hypersurface $\Sigma_U$.

In the following, we study the expansion of the Maxwell field $\mathcal{F}^{ab}$,  Killing vector $\chi^a$,  surface gravity $\kappa$,  energy-momentum stress  tensors $\mathcal{T}_{ab}^{\ae}$ and $\mathcal{T}_{ab}^{M}$, and Ricci tensor $\mathcal{R}_{ab}$.
The given source-free Maxwell field $\mathcal{F}^{ab}$ can be formulated in
terms of four-vectors representing physical fields. They are
 the electric field $E^a$ and  magnetic excitation $B^a$ as \cite{balakin},
 \begin{eqnarray}
E^a=\mathcal{F}^{ab}u_b, \quad B^a=\frac{e^{abmn}}{2\sqrt{-g}}\mathcal{F}_{mn}u_b,
\end{eqnarray}
where $e^{abmn}$ is the Levi-Civita tensor. For source free Maxwell field or from Eq.(\ref{motion}), it can be shown
$B^a=0$, that is, there is no need to consider $e^{abmn}$. Then, we find
\begin{eqnarray}
\mathcal{F}^{ab}=-E^au^b+E^bu^a.
\end{eqnarray}
On the other hand,  the electric field is spacelike, since
$E^au_a=0$. So, we have $E^a=(E\cdot s)s^a$. Thus,  $\mathcal{F}_{ab}=(E\cdot s)\varepsilon^{II}_{ab}$, where $\varepsilon^{II}_{ab}=(-s_au_b+s_bu_a)$.

Using Gauss law, there has \cite{bhattacharyya2}
\begin{eqnarray}
\nabla^b[F(r)\varepsilon_{ab}^{II}]=0\quad\Longrightarrow\quad F(r)=\frac{F_0}{b(r)^{n/2}},
\end{eqnarray}
where $F_0$ is a constant, $b(r)=r^2$. From Eq. (\ref{motion}),
 we can see $(E\cdot s)=\tilde{Q}/r^n$, where $\tilde{Q}$ is an integral constant, representing the total charge $Q$ of the space-time:
 \begin{eqnarray}
\label{}
Q=\frac{1}{4\pi}\int_{\infty}\frac{\tilde{Q}}{r^n}dA
=\frac{1}{4\pi}\int_{\infty}\tilde{Q}d\Omega_n=\frac{\Omega_n}{4\pi}\tilde{Q},
\end{eqnarray}
where $A$ and $\Omega_n$ are the area and the volume of $n$-sphere, $A=r^n\Omega_n$, $\Omega_n=2\pi^{d/2}/\Gamma(d/2)$, where $d=D-1$.
 Therefore, we have
\begin{eqnarray}
\label{Maxwellb}
\mathcal{F}_{ab}=\frac{\tilde{Q}}{r^n}(u_as_b-u_bs_a).
\end{eqnarray}

The Einstein,  aether  and Maxwell equations of motion (\ref{motion}) can be decomposed  by using the tetrad $u^a$, $s^a$ and $\hat{g}^{ab}$ defined above,
\begin{eqnarray}\label{projection}
&&\mathcal{T}^{\ae}_{ab}=\mathcal{T}^{\ae}_{uu}u_au_b-2\mathcal{T}^{\ae}_{us}u_{(a}s_{b)}
+\mathcal{T}^{\ae}_{ss}s_as_b+\frac{\hat{\mathcal{T}}_{\ae}}{n}\hat{g}_{ab},\nonumber\\
&&\mathcal{R}_{ab}=\mathcal{R}_{uu}u_au_b-2\mathcal{R}_{us}u_{(a}s_{b)}
+\mathcal{R}_{ss}s_as_b+\frac{\mathcal{\hat{R}}}{n}\hat{g}_{ab},\nonumber\\
&&\mathcal{T}^{M}_{ab}=\mathcal{T}^{M}_{uu}u_au_b-2\mathcal{T}^{M}_{us}u_{(a}s_{b)}
+\mathcal{T}^{M}_{ss}s_as_b+\frac{\hat{\mathcal{T}}_{M}}{n}\hat{g}_{ab}.
\end{eqnarray}
The coefficients of $\mathcal{T}^{\ae}_{ab}$ and $\mathcal{T}^{M}_{ab}$ in (\ref{projection}) can be computed from the general expression (\ref{EMTs}). The corresponding coefficients for $\mathcal{R}_{ab}$, on the other hand, are computed from the definition $[\nabla_a,~\nabla_b]X^c \equiv -\mathcal{R}^c_{~abd}X^d$ by choosing $X^a = u^a$ or $s^a$, and then contracting the resulting expressions again with $u^a$ and/or $s^a$ appropriately. The coefficients for the  three $(u, s)$ cross terms are
\begin{eqnarray}\label{usT}
\mathcal{T}^{\ae}_{us}=c_{14}\left[\hat{K}(a\cdot s)+\nabla_u(a\cdot s)\right],\quad
\mathcal{T}^{M}_{us}=0,\quad \mathcal{R}_{us}=(K_0-\frac{\hat{K}}{n})\hat{k}-\nabla_s\hat{K}.
\end{eqnarray}
The aether equation $s\cdot{\AE}=0$ and   the $us$-component  $\mathcal{R}_{us}=\mathcal{T}^{\ae}_{us} + 8\pi G_{\ae} \mathcal{T}^{M}_{us}$ yield
\begin{eqnarray}\label{usA}
 && c_{123}\nabla_s K_0+c_{13}(K_0-\frac{\hat{K}}{n})\hat{k}+c_2\nabla_s\hat{K}=\mathcal{T}^{\ae}_{us},\\
 \label{usB}
&&  c_{123}\nabla_s K-(1-c_{13})\mathcal{T}^{\ae}_{us} = 0.
\end{eqnarray}

After rewriting the motion Eq. (\ref{motion}) as
\begin{eqnarray}\label{motion2}
\mathcal{R}_{ab}-\frac{2}{n}\Lambda g_{ab}=\mathcal{T}^{\ae}_{ab}-\frac{1}{n}g_{ab}\mathcal{T}^{\ae}+8\pi G_{\ae}\big[\mathcal{T}^M_{ab}-\frac{1}{n}g_{ab}\mathcal{T}^M\big],
\end{eqnarray}
where $\mathcal{T}=g^{ab}\mathcal{T}_{ab}$, then the  $uu$- and $ss$-components of the gravitational field equations  give
\begin{eqnarray}\label{uussA}
\Big(1-\frac{n-1}{n}c_{14}\Big)\nabla\cdot a-(1-c_{13})K_{ab}K^{ab}+\frac{2}{n}\Lambda-\frac{c_{123}}{n}\nabla_c(Ku^c)-(1+c_2)\nabla_uK
+\frac{2(1-n)}{n}\frac{\tilde{Q}^2}{r^{2n}}=0,\\
\label{uussB}
(1-c_{13})\nabla_c(K_0u^c)+\frac{c_{123}}{n}(Ku^c)-\big[a^2+\frac{\hat{k}^2}{n}+\nabla_s(a\cdot s+\hat{k})+\frac{2}{n}\Lambda\big] \nonumber\\
-c_{14}(\frac{1}{n}\nabla\cdot a-a^2)+\frac{2(n-1)}{n}\frac{\tilde{Q}^2}{r^{2n}}=0.
\end{eqnarray}
In the next sections, we will use these equations to obtain two new  black holes solutions in the asymptotically flat spacetime in the case of $\Lambda=0$.

\section{Smarr formula}

In this section we  use Komar integral method to show that the procedure of deriving the Smarr formula in $D$-dimensional Einstein-Maxwell-aether theory.

To derive the Smarr formula, we first introduce the ADM mass (mass of a black hole), which is identical to the Komar mass defined in stationary spacetimes with the time translation Killing vector
$\chi^a$ \cite{berglund},
\begin{eqnarray}\label{KM}
M_{ADM} =-\frac{n}{8\pi G_{\ae}}\int_{{\cal{B}}_{\infty}}{\nabla^a\chi^b d\Sigma_{ab}},
\end{eqnarray}
where $d\Sigma_{ab} \equiv  - u_{[a}s_{b]}dA$, with $dA \; (\equiv r^nd\Omega_n)$ being the differential area  element  on the $n$-sphere ${\cal{B}}$,
and ${\cal{B}}_{\infty}$ is the sphere at infinity.

Now  we shall present the process of deriving Smarr formulas of the universal horizons for general $D$-dimensional  static and spherically symmetric Einstein-Maxwell-aether black holes. Let us first consider the geometric identity \cite{BCH},
\begin{eqnarray}
\label{ID}
\mathcal{R}_{ab}\chi^b=\nabla^b(\nabla_a\chi_b).
\end{eqnarray}
The derivative of the Killing vector $\chi^a=-(u\cdot\chi)u^a+(s\cdot\chi)s^a$ is given by
\begin{eqnarray}\label{kappa0}
\nabla^a\chi^b=-2\kappa u^{[a}s^{b]},\quad
\end{eqnarray}
where $\kappa$ denotes  the surface gravity usually defined in GR, and is given by
\begin{eqnarray}\label{kappa1}
\kappa =\sqrt{-\frac{1}{2}(\nabla_a\chi_b)(\nabla^a\chi^b)}=-(a\cdot s)(u\cdot\chi)+K_0(s\cdot\chi).
\end{eqnarray}
At the infinity, we have $(u\cdot\chi) = -1$ and $K_0(s\cdot\chi) = 0$. Then, Eq.(\ref{KM}) yields,
\begin{eqnarray}\label{KMb}
M_{ADM} =\lim_{r\rightarrow \infty}\left(\frac{n\Omega_nr^n(a\cdot s)}{8\pi G_{\ae}}\right).
\end{eqnarray}

Eq. (\ref{ID}) tell us that we can do an inner product of the right side of the Einstein field equation (\ref{motion2}) with the Killing vector $\chi^b$ as (for more detail see Appendix B)
\begin{eqnarray}
\label{IDa}
&& 8\pi G_{\ae}\Big (\mathcal{T}^M_{ab}-\frac{1}{n}g_{ab}\mathcal{T}^M\Big)\chi^b =\nabla^b\Big[\frac{2\tilde{Q}^2}{nr^{2n-1}}\; u_{[a}s_{b]}\Big],\nonumber\\
&&   \left(\mathcal{T}^{\ae}_{ab} - \frac{1}{n}g_{ab}\mathcal{T}^{\ae}\right)\chi^b = \nabla^b\Big\{\big[\big(\frac{c_{123}}{n} K  - c_{13} K_0\big) (s\cdot \chi)+(1-\frac{1}{n})c_{14}(a\cdot s)(u\cdot\chi)\big]u_{[a}s_{b]}\Big\} .
\end{eqnarray}
Then the inner product of Einstein field equation (\ref{motion2}) with $\chi^b$ can be cast in the form (in the case of $\Lambda=0$),
\begin{eqnarray}
\label{SmarrFa}
\nabla^bF_{ab}=0,\;\;\; F_{ab} \equiv 2\hat q(r)u_{[a}s_{b]},
\end{eqnarray}
where
\begin{eqnarray}
\label{SmarrFb}
&&\hat q(r)=q^M+q(r),\;\;q(r)= q^{\ae}+q^R,\;q^R\equiv\kappa=-(a\cdot s)(u\cdot\chi)+K_0(s\cdot\chi),\nonumber\\ &&q^{M}\equiv\frac{2\tilde{Q}^2}{nr^{2n-1}},\; q^{\ae}\equiv\big(\frac{c_{123}}{n} K  - c_{13} K_0\big) (s\cdot \chi)+(1-\frac{1}{n})c_{14}(a\cdot s)(u\cdot\chi).
\end{eqnarray}
On the other hand, comparing Eq.(\ref{SmarrFa}) with the soucre-free Maxwell equations  (\ref{motion}), we find that its solution must also take the form (\ref{Maxwellb}), that is,
$\hat q(r) =\hat q_0/r^n$.
Therefore, for asymptotically flat space-time we have,
\begin{eqnarray}
\label{ASYMP}
 (u\cdot\chi)\simeq -1,\;\;\;\; K_0(s\cdot\chi)\sim K(s\cdot\chi)\simeq 0,\;\;\;\;
 (a\cdot s) = \frac{\bar r_0}{2r^n} + {\cal{O}}(r^{-n-1}),
 \end{eqnarray}
 as $r\rightarrow\infty$, where $\bar r_0$ is a constant. Then, from Eq.(\ref{SmarrFb}) we find that $\hat q_0 = [1-(1-1/n)c_{14}]\bar r_0/2$. Thus, we have
\begin{eqnarray}
\hat q_\infty(r) =\left[1-\big(1-\frac{1}{n}\big)c_{14}\right]\frac{\bar r_0}{2r^n}.
\end{eqnarray}
Inserting Eq.(\ref{ASYMP}) into Eq.(\ref{KMb}),   we find that  the ADM mass  is given by \begin{eqnarray}
M_{ADM}=\frac{n\Omega_n\bar r_0}{16\pi G_{\ae}}.
\end{eqnarray}
And  the total mass $M$ of the spacetime is
\begin{eqnarray}\label{tmass}
M \equiv M_{ADM}+M_{\ae}  =\left[1-\big(1-\frac{1}{n}\big)c_{14}\right]\frac{n\Omega_n\bar r_0}{16\pi G_{\ae}}
=\frac{n}{8\pi G_{\ae}}\int_{\mathcal{B}_\infty}\hat qdA,
\end{eqnarray}
where $M_{\ae}=(1/n-1)c_{14}M_{ADM}$ is the aether mass or aether contribution to the renormalization of $M_{ADM}$.

On the other hand, using Gauss' law,  from Eq.(\ref{SmarrFa}) we find that
\begin{eqnarray}\label{tmassB}
   0 = \int_{\Sigma}{\left(\nabla_bF^{ab}\right)   d\Sigma_a} =   \int_{{\cal{B}}_{\infty}}{F^{ab}   d\Sigma_{ab}} - \int_{{\cal{B}}_{H}}{F^{ab}   d\Sigma_{ab}}
   =   \int_{{\cal{B}}_{\infty}}{\hat qdA} - \int_{{\cal{B}}_{H}}{\hat q dA}.
\end{eqnarray}
Here $d\Sigma_a$ is the surface element of a spacelike  hypersurface $\Sigma$, and $dA$ is the area element of $n$-dimensional sphere. The boundary $\partial\Sigma$ of $\Sigma$
consists of the boundary  at spatial infinity ${\cal{B}}_{\infty}$,  and the horizon ${\cal{B}}_{H}$, either the Killing or the universal.
Note that Eq.(\ref{tmassB}) is nothing but the conservation law of the flux of $F^{ab}$. Comparing the above expression   and
Eq.(\ref{SmarrFb}), we can construct the following Smarr formula in $D$-dimensions (similar to Ref. \cite{kastor}),
\begin{eqnarray}\label{smarrrelation}
(n-1)MG_{\ae}=\frac{nq_{H}A_{H}}{8\pi}+(n-1)V_{H}Q,
\end{eqnarray}
where $A_{H}$ are  the area of the universal or Killing horizon.
 The quantity $q_H$, like the surface gravity of the corresponding horizons, is the value of $q$ in (\ref{SmarrFb}) at the universal/Killing horizon. The electric potential is $V_H=4\pi Q/(n-1)\Omega_nr_H^{n-1}$.

In GR, from Eq. (\ref{kappa0}) and (\ref{ID}), the $q_H$ is just the surface gravity $\kappa_H$ on the usual Killing horizon. Now in the presence of aether field, it contains the aether contribution and becomes complicated.
However, the first law for the aether black hole may still be obtained via a variation of these Smarr relations for the uncharged case.
In the next section we consider it for  two new classes of exact charged aether black hole solutions.

For  the surface gravity at the universal horizon, when one  considers the peeling behavior of particles moving at any speed, i.e., capturing the role of the aether in the propagation of the physical rays, one finds that the  surface gravity at the universal horizon  is \cite{cropp,jacobson5,lin,greenwald}
\begin{eqnarray}\label{kappa}
\kappa_{UH}\equiv \frac{1}{2}\nabla_u(u\cdot\chi)
= \left.\frac{1}{2}\left(a\cdot s\right)\left(s\cdot \chi\right) \right|_{r= r_{UH}},
\end{eqnarray}
where in the last step we used the fact that $\chi_a$ is a Killing vector,  $\nabla_{(a}\chi_{b)} =0$.  It must be noted that this is different from the surface gravity defined in GR  by Eq.(\ref{kappa1}). In particular, at the universal horizon we have $u\cdot \chi = 0$, and Eq.(\ref{kappa1}) yields,
\begin{eqnarray}\label{kappa2}
\kappa\left(r_{UH}\right) = \left. K_0(s\cdot\chi)\right|_{r= r_{UH}}.
\end{eqnarray}

\section{Exact Solutions of charged aether black holes}

To construct exact solutions of charged aether black holes, let us first choose the Eddington-Finklestein coordinate system, in which
the    metric takes the form
\begin{eqnarray}\label{metric}
ds^2=-e(r)dv^2+2dvdr+r^2d\bar{\Omega}_n^2,
\end{eqnarray}
where  $d\bar\Omega_n^2=d\theta_1^2+\sin^2\theta_1d\theta_2^2+\cdots
+\sin^2\theta_1\cdots\sin^2\theta_{n-1}d\theta_n^2,\; (\theta_1, \theta_2, \cdots, \theta_{n-1}\in[0,\pi], \theta_n\in[0,2\pi])$ is the line element of $n$-sphere,.  And the corresponding timelike Killing  and aether vectors are
\begin{equation}
  \chi^a=(1,0,\cdots,0),\quad u^a=\big(\alpha,\beta,0,\cdots,0\big),\quad u_adx^a=\big(-e\alpha+\beta,\alpha,0,\cdots,0\big)\left(
\begin{array}{c}
 dv\\
 dr\\  d\theta_1\\ \cdots\\d\theta_n
\end{array}
\right),
\end{equation}
where $\alpha(r)$ and $\beta(r)$ are  functions of $r$ only. Then,  the metric can be written as  $g_{ab}=-u_au_b+s_as_b+\hat{g}_{ab}$, where we have the
 constraints $u^2=-1, ~s^2=1,~ u\cdot s=0$. The boundary condition of the aether components are such that
 \begin{eqnarray}
\label{aec}
\lim_{r\rightarrow\infty}u^a=(1,0,0,0),
\end{eqnarray}
and the metric coefficients are asymptotically flat.

Some quantities that explicitly appear in Eqs.(\ref{usA})-(\ref{uussB}) are  \cite{bhattacharyya2}
\begin{eqnarray}
\label{AS}
 (a\cdot s)=-(u\cdot \chi)',\quad K_0=-(s\cdot \chi)',\quad\hat{K}=-\frac{n(s\cdot \chi)}{r},\quad\hat{k}=-\frac{n(u\cdot \chi)}{r},
\end{eqnarray}
where a prime $(')$ denotes a derivative with respect to $r$. And $\alpha(r),~\beta(r)$ and $ e(r)$ are
\begin{eqnarray}\label{abe}
 \alpha(r)=\frac{1}{(s\cdot\chi)-(u\cdot\chi)},\quad \beta(r)=-(s\cdot \chi), \quad e(r)=(u\cdot\chi)^2-(s\cdot\chi)^2.
\end{eqnarray}
Then, from Eqs.(\ref{kappa}) and (\ref{kappa2}) we obtain
\begin{eqnarray}\label{abeA}
 \kappa_{UH} =-\left. \frac{1}{2}\left(u\cdot \chi\right)' (s\cdot\chi)\right|_{UH},\quad
  \kappa(r_{UH}) =-\left. \left(s\cdot \chi\right)' (s\cdot\chi)\right|_{UH}.
\end{eqnarray}
Clearly, in general $ \kappa_{UH} \not= \kappa(r_{UH})$.

Substituting Eq.(\ref{AS}) into (\ref{usT}), a straightforward  calculation yields
 \begin{eqnarray}
 \mathcal{R}_{us}=0.
\end{eqnarray}
So that the $us$-component gravitational field motion equation  $\mathcal{R}_{us}=\mathcal{T}^{\ae}_{us} + 8\pi G_{\ae} \mathcal{T}^{M}_{us}$ yields
\begin{eqnarray}
\mathcal{T}^{\ae}_{us}=0,
\end{eqnarray}
and the aether field motion Eq.(\ref{usB}) gives
\begin{eqnarray}
c_{123}\nabla_sK=0.
\end{eqnarray}
They both together with Eqs. (\ref{usT}) and (\ref{AS}) lastly give
\begin{eqnarray}\label{aetherEq}
&&\mathcal{T}^{\ae}_{us}=c_{14}\frac{(s\cdot\chi)}{r^n}[r^n(u\cdot\chi)']'=0,
\nonumber\\
&&c_{123}\nabla_sK=c_{123}(u\cdot\chi)\Big\{\frac{1}{r^n}\big[r^n(s\cdot\chi)
\big]'\Big\}'=0.
\end{eqnarray}
It is easy to see that there are many ways for satisfying these two equations, in the following,  we shall consider only two special cases $c_{14}=0,~c_{123}\neq0$ and $c_{123}=0,~c_{14}\neq0$ to obtain both kinds of exact solutions in asymptotically flat spacetime. According to Eq. (\ref{spin0}), the spin-0 speed diverges as the first case $c_{14}=0$, while it vanishes as the second case $c_{123}=0$. In the context of a black hole solution, when $c_{14}=0$, the spin-0 horizon coincides with the universal horizon; when $c_{123}=0$, it is pushed  away to the asymptotic boundary.

\subsection{The first kind aether black hole solutions for $c_{14}=0$}

When the coupling constant $c_{14}$ is set to zero and $c_{123}\neq 0$, from Eqs.(\ref{usB}) and (\ref{aetherEq}) one can see the quantity $\nabla_sK$ has to be vanished, i.e., $\nabla_sK=0$. So, the trace of the extrinsic curvature $K$ of the $\Sigma_U$ hypersurface is a  constant and, Eq. (\ref{aetherEq}) gives $(s\cdot\chi)=t_1r/(n+1)+t_2/r^n$, where $t_1,\;t_2$ are integral constants. From the condition (\ref{aec}), the constant $t_1$ should be zero, then it can be rewritten as
 \begin{eqnarray}\label{sx}
 &&(s\cdot\chi)=\frac{r_{\ae}^n}{r^n},
\end{eqnarray}
where $r_{\ae}$ is another constant.
Since a vanishing $(s\cdot\chi)$ signifies of the aether $u^a$ aligning with Killing vector $\chi^a$, so this constant measure the misalignment of the aether.
   Substituting Eq.(\ref{sx}) into (\ref{uussA}) or (\ref{uussB}), we obtain
   \begin{eqnarray}\label{aether}
 &&(u\cdot\chi)=-\sqrt{1-\frac{\bar{r}_0}{r^{n-1}}
 +\frac{\bar{Q}^2}{r^{2n-2}}+(1-c_{13})\frac{r^{2n}_{\ae}}{r^{2n}}}\;,\bar{r}_0=\frac{8\pi r_0}{n\Omega_n},\bar{Q}^2=\frac{2\tilde{Q}^2}{n(n-1)},\\
 \label{metric1}&&e(r)=1-\frac{\bar {r}_0}{r^{n-1}}+\frac{\bar{Q}^2}{r^{2n-2}}-c_{13}\frac{r^{2n}_{\ae}}{r^{2n}},
\end{eqnarray}
where $r_0$ is an integral constant which then shows the ADM mass of the black hole.
 If $c_{13}=0$, it is the usual $D$-dimensional Reissner-Nordstr\"{o}m black hole, but now with an universal horizon, which is very interesting.

The location of the universal horizon $r_{UH}$ is the root of equation $(u\cdot\chi)=0$ or $(u\cdot\chi)^2=0$ (it is similar to the location of the Killing horizon which is the largest root of the equation $\chi^2=0$). Meanwhile, $(u\cdot\chi)$ is a physical component of the aether, and should be regular and real everywhere. So,  from Eq.(\ref{aether}) one can see that the quantity $(u\cdot\chi)^2$ should be non-negative, i.e., $(u\cdot\chi)^2\geq0$. Therefore, the location of $r_{UH}$ is the minimum point of the function $(u\cdot\chi)^2$. It implies that $(u\cdot\chi)^2|_{r=r_{UH}}=0$ and $d(u\cdot\chi)^2/dr|_{r=r_{UH}}=0$. Then, $r_{\ae}$ becomes a function of $r_0$. That is, the global existence of the
aether reduces the number of three independent constants ($r_0, r_{\ae}, Q$) to two,  ($r_0, Q$).  Thus,  from the equations  $(u\cdot\chi)^2 = 0$ and $d(u\cdot\chi)^2/dr = 0$, we find
\begin{eqnarray}\label{ruh1}
r^{2n}_{\ae}=\frac{n-1}{1-c_{13}}\left[r_{UH}^{2n}
-\frac{\bar{r}_0}{2}r_{UH}^{n+1}\right],\quad
 r_{UH}^{n-1}=\frac{n+1}{4n}\bar{r}_0\left[1+
 \sqrt{1-\frac{16n\bar{Q}^2}{(n+1)^2\bar{r}_0^2}}\right].
\end{eqnarray}
One can see that the charge $Q$ is subjected to the condition
$ \bar Q\leq(n+1)\bar r_0/4\sqrt{n}$, in order to have $r_{UH}$ real.
When $\bar Q=\bar r_0/2$, we find $r^{2n}_{\ae}=0$ and $r^{n-1}_{UH}=\bar r_0/2$. When $\bar Q>\bar r_0/2$, we have $r_{\ae}^{2n}<0$. Thus, in order to have the aether be regular everywhere, the charge should be,
\begin{eqnarray}
 \bar Q\leq\frac{\bar r_0}{2},
\end{eqnarray}
which is the same as  that given in the Reissner-Nordstrom black hole.

Now let us derive the Smarr formula and the effective first law at the universal horizon.  The surface gravity at the universal horizon can be computed via (\ref{kappa}) and given by
\begin{eqnarray}\label{kappauh1}
 \kappa_{UH}=\frac{1}{2}\nabla_u(u\cdot\chi)|_{r_{UH}}=\frac{n-1}{2\sqrt{1-c_{13}}r_{UH}}
\sqrt{\frac{n}{n+1}\Big(
 1-\frac{\bar Q^2}{r_{UH}^{2n-2}}\Big)\Big(
 1-\frac{\bar Q^2}{nr_{UH}^{2n-2}}\Big)}.
\end{eqnarray}
If one uses definition (\ref{kappa2}), then this surface gravity is
\begin{eqnarray}\label{kappauh11}
 \kappa(r_{UH})=K_0(s\cdot\chi)|_{r_{UH}}=\frac{n(n-1)}{(n+1)(1-c_{13})r_{UH}}\Big(
 1-\frac{\bar Q^2}{\;r_{UH}^{2n-2}}\Big),
\end{eqnarray}
which is different.
And the Eq. (\ref{SmarrFb}) becomes
\begin{eqnarray}\label{}
 q_{UH}=\frac{n(n-1)}{(n+1)r_{UH}}\Big(
 1-\frac{\bar Q^2}{\;r_{UH}^{2n-2}}\Big),\;q^M_{UH}=\frac{(n-1)\bar Q^2}{r^{2n-1}_{UH}}.
\end{eqnarray}
From (\ref{SmarrFb}) and (\ref{tmass}), one get the Smarr formula
\begin{eqnarray}\label{smarr1}
(n-1)MG_{\ae}=\frac{n}{8\pi} q_{UH}A_{UH}+(n-1)V_{UH}Q,\;V_{UH}=\frac{4\pi Q}{(n-1)\Omega_nr^{n-1}_{UH}}.
\end{eqnarray}
After variation it with respect to $A_{UH}$ and $Q$, one get the differential form
\begin{eqnarray}\label{first}
G_{\ae}\delta M=\frac{n-1}{(n+1)8\pi r_{UH}} \Big(n-\frac{\bar Q^2}{r_{UH}^{2n-2}}\Big)\delta A_{UH}+\frac{2}{n+1}V_{UH}\delta Q,
\end{eqnarray}
which might be used as the effective first law to these Einstein-aether black holes. One can see that the prefactor of $\delta A_{UH}$ isn't proportional either to $\kappa(r_{UH})$ or $\kappa_{UH}$!
In the following, we study thermodynamic properties of the universal horizons in three cases.
\subsubsection{Neutral case $Q=0$ and $c_{13}=0$}
When the charge $Q\rightarrow0$ and $c_{13}=0$, the metric function (\ref{metric1}) becomes
\begin{eqnarray}\label{}
e(r)=1-\frac{\bar r_0}{r^{n-1}},
\end{eqnarray}
 which is just the $D$-dimensional Schwarzschild black hole with an universal horizon inside. Its universal horizon and Killing horizon are
\begin{eqnarray}\label{}
r_{UH}^{n-1}=\frac{n+1}{2n}\bar r_0,\; r_{KH}^{n-1}=\bar r_0.
\end{eqnarray}
  Now the surface gravity at the universal horizon are given by
\begin{eqnarray}\label{kappauhq}
&& \kappa_{UH}=\frac{n-1}{2r_{UH}}
\sqrt{\frac{n}{n+1}},\\&&
 \kappa(r_{UH})=\frac{n-1}{r_{UH}}\frac{n}{n+1}.
\end{eqnarray}
The Smarr formula and first law can be constructed in this case,
\begin{eqnarray}\label{}
&& (n-1)MG_{\ae}=nTS,\\&&
 \delta MG_{\ae}=T\delta S,\label{firstlaw}
\end{eqnarray}
where the temperature can be defined as $T=\sqrt{n/(n+1)}\kappa_{UH}/\pi$ or $T=\kappa(r_{UH})/2\pi$ and, the entropy as $S=A_{UH}/4$.
\subsubsection{Charged case $Q\neq0$ but $2\bar Q/\bar r_0\ll1$}

In this case, the surface gravity at the universal horizon are given by
\begin{eqnarray}\label{}
&& \kappa_{UH}=\frac{n-1}{2\sqrt{1-c_{13}}r_{UH}}
\sqrt{\frac{n}{n+1}},\\&&
 \kappa(r_{UH})=\frac{n-1}{(1-c_{13})r_{UH}}\frac{n}{n+1}.
\end{eqnarray}
The Smarr formula and first law can be constructed in this case,
\begin{eqnarray}\label{}
&& (n-1)MG_{\ae}=nTS+(n-1)V_{UH}Q,\\&&
 \delta MG_{\ae}=T\delta S+\frac{2}{n+1}V_{UH}\delta Q,\label{}
\end{eqnarray}
where the temperature can be defined as $T=\sqrt{n(1-c_{13})/(n+1)}\kappa_{UH}/\pi$ or $T=(1-c_{13})\kappa(r_{UH})/2\pi$ and, the entropy as $S=A_{UH}/4$. It is easy to see that one can still construct the first law which is slightly modified.
\subsubsection{Extremely high dimension case $n\rightarrow\infty$ and $c_{13}=0$}
When  $n\rightarrow\infty$ and $c_{13}=0$,
the metric function (\ref{metric1}) becomes
\begin{eqnarray}\label{}
e(r)=1-\frac{\bar r_0}{r^{n-1}}+\frac{\bar Q^2}{r^{2n-2}},
\end{eqnarray}
 which is just the $D$-dimensional Reissner-Nordstr\"{o}m black hole with an universal horizon inside. In this case, the location of universal and Killing horizons are give by
  \begin{eqnarray}r^{n-1}_{UH}\rightarrow\bar r_0/2,\;r_{KH}^{n-1}=\frac{1}{2}\big(\bar r_0\pm \sqrt{\bar r_0^2-4\bar Q^2}\;\big).
 \end{eqnarray}
 Now the surface gravity at the universal horizon are given by
\begin{eqnarray}\label{}
&& \kappa_{UH}=\frac{n-1}{2r_{UH}}
\sqrt{ 1-\frac{\bar Q^2}{r_{UH}^{2n-2}}},\\&&
 \kappa(r_{UH})=\frac{n-1}{r_{UH}}\Big(
 1-\frac{\bar Q^2}{r_{UH}^{2n-2}}\Big).\label{t2}
\end{eqnarray}
The Eqs. (\ref{smarr1}) and (\ref{first}) becomes
\begin{eqnarray}\label{}
&&(n-1)MG_{\ae}=\frac{n\kappa(r_{UH})}{8\pi} A_{UH}+(n-1)V_{UH}Q,\\
&&\delta MG_{\ae}=\frac{n-1}{8\pi r_{UH}}\delta A_{UH},
\end{eqnarray}
where the work term $V_{UH}\delta Q$ vanishes, like the Eq. (\ref{first2}) of the second kind aether black hole. By Comparing it to Eq. (\ref{firstlaw}), one can get the temperature as
 \begin{eqnarray}\label{}
 T=\frac{n-1}{2\pi r_{UH}},
\end{eqnarray}
which also can be obtained by dividing Eq. (\ref{t2}) by $2\pi$ and letting $2\bar Q/\bar r_0\ll1$.
 So one can see that in the much higher dimension, like the second kind aether black hole, the electric charge should be constrained severely.
\subsubsection{Extremely charged case $\bar Q\rightarrow \bar r_0/2$ and $c_{13}=0$}
In this case, the metric function (\ref{metric1}) becomes
\begin{eqnarray}\label{}
e(r)=\Big(1-\frac{\bar r_0}{2r^{n-1}}\Big)^2,
\end{eqnarray}
which is just the extremal $D$-dimensional Reissner-Nordstr\"{o}m black hole.
Its universal and Killing horizons are given by
\begin{eqnarray}\label{}
r^{n-1}_{UH}=r^{n-1}_{KH}=\frac{\bar r_0}{2},
\end{eqnarray}
which shows that the universal horizon is naked to outside.
 Now the surface gravity at the universal horizon are given by
\begin{eqnarray}\label{}
&& \kappa_{UH}= \kappa(r_{UH})=0,
\end{eqnarray}
which means the corresponding temperature is zero and the thermal description breaks down.
The Eq. (\ref{smarr1}) becomes
\begin{eqnarray}\label{}
MG_{\ae}=V_{UH}Q=\sqrt{\frac{n}{2(n-1)}}Q.
\end{eqnarray}

\subsection{The second kind aether black hole solutions for $c_{123}=0$}

When the coupling constant $c_{123}$ is set to zero and $c_{14}\neq 0$,   Eq. (\ref{aetherEq}) gives
 \begin{eqnarray}\label{ux}
 &&(u\cdot\chi)=-1+\frac{\bar{r}_0}{2r^{n-1}},
\end{eqnarray}
where $\bar r_0$ is  a constant related to the black holes' mass.
   Substituting it into  Eq. (\ref{uussA}) or (\ref{uussB}), we obtain
\begin{eqnarray}
 &&(s\cdot\chi)=\frac{\bar{r}_0+2r_u}{2r^{n-1}},\;
 e(r)=1-\frac{\bar{r}_0}{r^{n-1}}-\frac{r_u(r_u+\bar{r}_0)}{r^{2n-2}},\label{metric2}\\
 &&r_u=\frac{\bar{r}_0}{2}\Big[\sqrt{J-\frac{4\bar Q^2}{(1-c_{13})\bar r_0^2}+1}-1\Big],\;J=\frac{c_{13}-(1-1/n)c_{14}}{1-c_{13}}.
\end{eqnarray}
From the above expressions, we find
\begin{eqnarray}
 \alpha(r)=\frac{1}{(s\cdot\chi)-(u\cdot\chi)}=\frac{1}{1+\frac{r_u}{r^{n-1}}}.
\end{eqnarray}
Since  it is one of the component of $u^a$, it should be regular everywhere (possibly except at the singular point $r = 0$), we must have
\begin{eqnarray}\label{ru}
r_u\geq0\Rightarrow \bar Q\leq \sqrt{(1-c_{13})J}\; \frac{\bar r_0}{2}\;\;,\quad J\geq 0.
\end{eqnarray}
Note that the constraint $J \ge 0$ on the coupling constants means that \begin{eqnarray}
 c_{14} \le \frac{n}{n-1}c_{13},
\end{eqnarray}
 which is stronger than (\ref{CDs}). It is also means that the charge should be constrained severely small from Eq. (\ref{ru}). The gravitational wave event GW170817 and the one of gamma-ray burst GRB170817 provide much severe constrain that $c_{13}<10^{-15}$ \cite{oost}. Therefore one can easily get the condition $2\bar Q/\bar r_0\ll1$ for the second kind aether black hole.

The position of the universal horizon $r_{UH}$ and its surface  gravity (using Eq. \ref{kappa}) are
\begin{eqnarray}\label{kappauh2}
 r_{UH}^{n-1}=\frac{\bar r_0}{2},\quad  \kappa_{UH}=\frac{n-1}{2r_{UH}}
\sqrt{J-\frac{\bar Q^2}{(1-c_{13})r_{UH}^{2n-2}}+1}\;.
\end{eqnarray}
If one uses definition (\ref{kappa2}), then this surface gravity is
\begin{eqnarray}\label{}
  \kappa(r_{UH})=\frac{n-1}{r_{UH}}
\Big[J-\frac{\bar Q^2}{(1-c_{13})r_{UH}^{2n-2}}+1\Big].
\end{eqnarray}
And the Eq. (\ref{SmarrFb}) becomes
\begin{eqnarray}\label{}
 q_{UH}=\frac{n-1}{r_{UH}}
\Big[1-(1-\frac{1}{n})c_{14}-\frac{\bar Q^2}{r_{UH}^{2n-2}}\Big],\;q_{UH}^M=\frac{(n-1)\bar Q^2}{r^{2n-1}_{UH}}.
\end{eqnarray}

From (\ref{SmarrFb}) and (\ref{tmass}), one get the Smarr formula
\begin{eqnarray}\label{smarr2}
(n-1)MG_{\ae}&=&\frac{n}{8\pi} (q_{UH}+q_{UH}^M)A_{UH}\nonumber\\
&=&\frac{n}{8\pi} q_{UH}A_{UH}+(n-1)V_{UH}Q.
\end{eqnarray}
One can rewritten Eq. (\ref{smarr2}) as
\begin{eqnarray}\label{}
(n-1)MG_{\ae}&=&\frac{n}{8\pi} (q_{UH}+q_{UH}^M)A_{UH}=(n-1)\Big[1-(1-\frac{1}{n})c_{14}\Big]\frac{nA_{UH}}{8\pi r_{UH}},
\end{eqnarray}
which involves no $Q$ term! After varying it with respect to $A_{UH}$, one can get the differential form
\begin{eqnarray}
G_{\ae}\delta M=\Big[1-(1-\frac{1}{n})c_{14}\Big]\frac{n-1}{8\pi r_{UH}}\delta A_{UH}.\label{first2}
\end{eqnarray}
So that there is no term involving charge $Q$.
In the following, we study thermodynamic properties of the universal horizons in two specific cases.

\subsubsection{Neutral case $Q\rightarrow0$}
The metric function (\ref{metric2}) can be rewritten as
\begin{eqnarray}\label{}
e(r)=1-\frac{\bar r_0}{r^{n-1}}-\frac{J\bar r_0^2}{4r^{2n-2}}+\frac{\bar Q^2}{(1-c_{13})r^{2n-2}}.
\end{eqnarray}
When the charge $Q\rightarrow0$,
it becomes
\begin{eqnarray}\label{}
e(r)=1-\frac{\bar r_0}{r^{n-1}}-\frac{J\bar r_0^2}{4r^{2n-2}},
\end{eqnarray}
and its universal horizon and Killing horizon are given by
\begin{eqnarray}\label{}
r^{n-1}_{UH}=\frac{\bar r_0}{2},\;r^{n-1}_{KH}=\frac{\bar r_0}{2}(1+\sqrt{1+J}).
\end{eqnarray}
The surface gravity at the universal horizon are given by
\begin{eqnarray}\label{kappauhs}
&& \kappa_{UH}=\frac{n-1}{2r_{UH}}\sqrt{J+1}\;,\\&&
 \kappa(r_{UH})=\frac{n-1}{r_{UH}}(J+1).
\end{eqnarray}
The Smarr formula and first law of this case are
\begin{eqnarray}\label{}
&& (n-1)MG_{\ae}=(1-c_{13})nTS,\\&&
 \delta MG_{\ae}=(1-c_{13})T\delta S,
\end{eqnarray}
where the temperature can be defined as $T=\sqrt{J+1}\kappa(r_{UH})/\pi$ or $T=\kappa(r_{UH})/2\pi$ and, the entropy as $S=A_{UH}/4$.

\subsubsection{Extremely charged case $\bar Q\rightarrow\sqrt{(1-c_{13})J}\;\bar r_0/2$}
When the charge is extremely large, i.e., from Eq. (\ref{ru}), there has $r_u\rightarrow0$, then the metric function becomes
\begin{eqnarray}
 e(r)=1-\frac{\bar{r}_0}{r^{n-1}},
\end{eqnarray}
which is just $D$-dimensional Schwarzschild black hole but now it is charged. Its universal horizon and Killing horizon are given by
\begin{eqnarray}\label{}
r^{n-1}_{UH}=\frac{\bar r_0}{2},\;r^{n-1}_{KH}=\bar r_0.
\end{eqnarray}
The surface gravity at the universal horizon are given by
\begin{eqnarray}\label{kappauhs}
&& \kappa_{UH}=\frac{n-1}{2r_{UH}},\\&&
 \kappa(r_{UH})=\frac{n-1}{r_{UH}}.
\end{eqnarray}
The Eqs. (\ref{smarr2}) and (\ref{first2}) become
\begin{eqnarray}\label{}
&&(n-1)MG_{\ae}=(1-c_{13})\frac{n\kappa(r_{UH})}{8\pi}A_{UH}+(n-1)V_{UH}Q,\\&&
G_{\ae}\delta M=\Big(1-c_{13}+\frac{4\bar Q^2}{\bar r_0^2}\Big)\frac{\kappa(r_{UH})}{8\pi}\delta A_{UH}.
\end{eqnarray}
If using the condition $2\bar Q/\bar r_0\ll1$, one can obtain that
\begin{eqnarray}\label{}
&&(n-1)MG_{\ae}=(1-c_{13})nTS,\\&&
G_{\ae}\delta M=(1-c_{13})T\delta S,
\end{eqnarray}
which shows that the first law of black hole thermodynamics holds, and the temperature $T=\kappa_{UH}/\pi$ or $T=\kappa(r_{UH})/2\pi$, the entropy $S=A_{UH}/4$.

\section{Conclusions and discussion}

In this paper, we have studied the static, spherically symmetric, asymptotically flat black hole solutions of Einstein-Maxwell-aether theory and the linearized wave spectrum in the higher $D$-dimensional spacetime. We present two new kinds of charged $D$-dimensional black holes with universal horizons: the first kind $c_{14}=0, c_{123}\neq0$; the second kind $c_{14}\neq0, c_{123}=0$. In both of cases, the universal horizons are independent of the coupling constants $c_i$. It is interesting that there exist  $D$-dimensional Schwarzschild, Reissner-Nordstr\"{o}m and charged Schwarzschild black holes but now with an universal horizon inside.

We first study the Smarr relation associated with the universal horizon like the usual Killing horizon in GR. With the use of ADM mass, we find the total mass (\ref{tmass}) of an asymptotically flat solution defined in the asymptotic aether rest frame. Then the Smarr relation (\ref{smarrrelation}) is constructed with the total mass. One will presume the quantity $q_{UH}$ is identical to the surface gravity at the universal horizon. However, there are two definitions (\ref{kappa}) and (\ref{kappa2}). It is proportional to $\kappa(r_{UH})$, i.e., $q_{UH}=(1-c_{13})\kappa(r_{UH})$, but not proportional to $\kappa_{UH}$. So the definition (\ref{kappa2}) seems pleasant. But the definition (\ref{kappa}) captures the role of the aether in the propagation of the physical rays and is more reasonable. For the charged aether black holes, they are completely different from each other. In order to make them consistent with each other, we have to set the charge $Q=0$ or $2\bar Q/\bar r_0\ll1$. Only in this way, both definitions  have slightly difference.

We then consider the first law of a black hole mechanics associated with the universal horizon. We variate the obtained Smarr relation with respect to the universal horizon area and charge and find that: for the both kinds charged aether black holes, the prefactor of $\delta A_{UH}$ aren't proportional either to $\kappa(r_{UH})$ or $\kappa_{UH}$. For the second kind black hole and the first kind black hole in the case of extreme high dimension, there is no work term involving charge $Q$.  We consider some special cases: neutral, near neutral and extremely charged. For the neutral aether black hole, both kinds of them can have a slightly modified first law of black hole mechanics associated with the universal horizon. For the charged but $2\bar Q/\bar r_0\ll1$ first kind aether black hole, a slightly modified first law can still be constructed.

We lastly study the linearized aether wave spectrum and find that there are a total of five wave modes. The first two are the usual spin-2 gravitational waves---transverse $\epsilon_{jl}\neq0$ and traceless $\epsilon_{aa}=0,\epsilon_I=\epsilon_{Id}=0$ whose speeds are the same as those in 4-dimensional spacetime. The second two modes are spin-1 transverse aether wave $\epsilon_I\neq0,$ and $\epsilon_{Id}\neq0$ whose speeds are also the same as those in $D=4$ spacetime. The fifth mode is spin-0 trace aether wave---nonzero $\epsilon_{00},\;\epsilon_{dd},\;\epsilon_{II}$ whose speed is dependant on dimension number $n$.

Note that in the case of the cosmological constant $\Lambda\neq0$, we here also give both exact solutions and construct topological charged Einstein-aether (anti) de Sitter black holes.
In the case of $(c_{14}=0,~c_{123}\neq0)$, from Eq. (\ref{aetherEq}), one can get $(s\cdot\chi)=r_{\ae}^{n}/r^{n}+\sqrt{\Lambda'}r/(n+1)$, where $\Lambda'$ is a constant.
Then the exact solution is
\begin{eqnarray}
  e(r)=
  1-\frac{\bar r_0+2r_{\ae}^nl_s}{r^{n-1}}+\frac{\bar Q^2}{r^{2n-2}}-c_{13}\frac{r_{\ae}^{2n}}{r^{2n}}
 +[(c_2+c_{123}/n)l_s^2-\bar\Lambda] r^2,
\end{eqnarray}
which is similar to that in ref. \cite{lin2017} when $k=1$, where $l_s=\sqrt{\Lambda'}/(n+1)$, $\bar \Lambda=2\Lambda/n(n+1)$. In the case of $(c_{14}\neq0,~c_{123}=0)$, it is
\begin{eqnarray}
  e(r)=1-\frac{\bar r_0}{r^{n-1}}+\frac{(n-1)c_{14}-nc_{13}}{n(1-c_{13})}\frac{\bar r_0^2}{4r^{2n-2}}
 +\frac{1}{1-c_{13}}(\frac{\bar Q^2}{r^{2n-2}}-\bar\Lambda r^2),
\end{eqnarray}
which is also similar to that in ref. \cite{lin2017} when $k=1$.
The topological charged
Einstein-aether (anti) de Sitter  black holes are
\begin{equation}
ds^2 = -\big[e(r) - 1\big] dv^2 + 2dv dr + r^2d\bar\Omega_n^2.
\end{equation}
Study of their properties is out of the scope of this paper and let us consider them in the near future.

\begin{acknowledgments} The authors would like to thank Prof. Ted Jacobson for timely help. Ding's work is supported by the National Natural Science Foundation (NNSFC)
of China (grant No. 11247013), Hunan Provincial Natural Science Foundation of China (grant No. 2015JJ2085), and the fund under grant No. QSQC1708; Wang's work is supported  by NNSFC of China (Nos.
11375153, 11675145).
\end{acknowledgments}

\appendix
\section{Einstein-aether wave}
In this section, we study the linearized Einstein-aether theory in higher $D$-dimensional spacetime and find the speeds and polarizations of all the wave modes. The first step is to linearize the field equations about the flat background solution with Minkowski metric and constant unit vector $\b u^a$. The fields are expanded as
\begin{eqnarray}
g_{ab}=\eta_{ab}+\gamma_{ab},\;u^a=\b u^a+v^a,
\end{eqnarray}
where $\eta_{ab}=$diag$(-1,1,\cdots,1)$  for $D$-dimensional flat Minkowski spacetime $(x^0,x^i)$ and $\b u^a=(1,0,\cdots,0), (i=1,2,\cdots,D-1)$.

Keeping only the first order terms in $v^a$ and $\gamma_{ab}$, the field equations (\ref{motion}) become (set $\Lambda=0, Q=0$)
\begin{eqnarray}\label{line}
\mathcal{G}_{ab}^{(1)}=\mathcal{T}_{ab}^{\ae(1)},\;\;\partial_aY^{(1)a}_{~~~~~m}=-\lambda \b u_m,\;\;v^0-\frac{1}{2}\gamma_{00}=0,
\end{eqnarray}
where the superscript (1) denotes the first order part of the corresponding quantity, and
\begin{eqnarray}\label{}
&&\mathcal{T}_{ab}^{\ae(1)}=\partial_0Y_{(ab)}^{(1)}-\partial_mY_{~~(a}^{(1)m}\b u_{b)},\nonumber\\&&
\mathcal{G}_{ab}^{(1)}=\frac{1}{2}\Box\gamma_{ab}
+\frac{1}{2}\gamma_{,ab}-\gamma_{m(a,b)}^{\quad\quad m}-\frac{1}{2}\eta _{ab}(\Box\gamma-\gamma_{mn,}^{\quad mn}),
\end{eqnarray}
where $\gamma=\gamma^m_m$. The linearized quantity $Y^{(1)}_{ab}$ is given by
\begin{eqnarray}\label{}
Y_{ab}^{(1)}=c_1\nabla_au_b +c_2\eta_{ab}\nabla_mu^m+c_3\nabla_bu_a-c_4\b u_a\nabla_0u_b,
\end{eqnarray}
where the quantity $\nabla_au_b$ are expanded to linear order, and replaced by
\begin{eqnarray}\label{}
(\nabla_au_b)^{(1)}=v_{b,a}+\frac{1}{2}(\gamma_{ab,0}+\gamma_{0b,a}-\gamma_{a0,b}).
\end{eqnarray}

Applying the gauge conditions \cite{jacobson4}
$
\gamma_{0i}=0,\;v_{i,i}=0
$, one obtains the tensors
\begin{eqnarray}\label{}
&&Y^{(1)a}_{ai,}=-c_{14}(v_{i,00}-\frac{1}{2}\gamma_{00,i0})+c_1v_{i,kk}+\frac{1}{2}c_{13}\gamma_{ik,k0}
+\frac{1}{2}c_2\gamma_{kk,0i},\\&&
\mathcal{G}^{(1)}_{ij}=\frac{1}{2}\Box\gamma_{ij}+\frac{1}{2}\gamma_{,ij}
+\gamma_{k(i,j)k}+\frac{1}{2}\delta_{ij}(\Box\gamma-\gamma_{00,00}-\gamma_{kl,kl}),\\&&
\mathcal{T}^{\ae(1)}_{ij}=c_{13}(v_{(i,j)0}+\frac{1}{2}\gamma_{ij,00})
+\frac{1}{2}c_2\delta_{ij}\gamma_{kk,00},
\end{eqnarray}
where $i,j,k,l$ denotes spatial coordinate indices and repeated ones are summed with the Kronecker delta.

The perturbations are of the plane wave form
\begin{eqnarray}\label{ans}
\gamma_{ab}=\epsilon_{ab}e^{ik_cx^c},\;v^a=\epsilon^ae^{ik_cx^c},
\end{eqnarray}
and the coordinates are chosen such that the wave vector is $(k_0,0,\cdots,0,k_d)$, where $d=D-1$.
The gauge conditions $\gamma_{0i}=0,\;v_{i,i}=0$ gives
\begin{eqnarray}
\epsilon_{0i}=0,\;\epsilon_d=0.
\end{eqnarray}
Substituting the plane wave ansatz (\ref{ans}) into the field equations (\ref{line}), one can obtain
\begin{eqnarray}
&&[A_I]~~~(c_{14}s^2-c_1)\epsilon_I-\frac{1}{2}c_{13}\epsilon_{Id}=0,\label{AI} \\
&&[A_d]~~~c_{14}\epsilon_{00}+c_{123}\epsilon_{dd}+c_2\epsilon_{II}=0, \label{Ad}\\
&&[E_{II}]~~~\epsilon_{00}+(1+c_2)s^2\epsilon_{dd}
+\frac{n-1}{n}\Big[(1+c_2+\frac{c_{123}}{n-1})s^2-1\Big]\epsilon_{II}=0,\label{EII}\\
&&[E_{jl}]~~~[(1-c_{13})s^2-1]\epsilon_{jl}=0, \label{Ejl}\\
&&[E_{Id}]~~~c_{13}\epsilon_{I}+(c_{13}-1)s\;\epsilon_{Id}=0,\label{EId}\\
&&[E_{dd}]~~~c_{123}\epsilon_{dd}+(1+c_2)\epsilon_{II}=0,\label{Edd}
\end{eqnarray}
where $j\neq l$ and $j,l,I=1,2,\cdots,n$, $n=D-2$. $[A_i]$ and $[E_{ij}]$ indicate the components of the aether and Einstein equations, $s=k_0/k_d$ for the wave speed, $\epsilon_{II}=\epsilon_{11}+\epsilon_{22}+\cdots+\epsilon_{nn}$ is the trace of the transverse spatial part of the metric polarization $\epsilon_{ab}$. Another equation is
\begin{eqnarray}\label{E22}
&&[E_{11}-E_{22}+\cdots+E_{(n-1)(n-1)}-E_{nn}]\nonumber\\ &&[(1-c_{13})s^2-1]
(\epsilon_{11}-\epsilon_{22}+\cdots+\epsilon_{(n-1)(n-1)}-\epsilon_{nn})=0
\end{eqnarray}
for even $n$, or
\begin{eqnarray}\label{E221}
&&[E_{11}-\frac{E_{22}+E_{33}}{2}+\cdots+E_{(n-1)(n-1)}-E_{nn}]\nonumber\\ &&[(1-c_{13})s^2-1]
(\epsilon_{11}-\frac{\epsilon_{22}+\epsilon_{33}}{2}+\cdots+\epsilon_{(n-1)(n-1)}-\epsilon_{nn})=0
\end{eqnarray}
for odd $n$.

There are a total of five modes. The first two modes corresponding to the usual gravitational waves in GR are found when polarization components $\epsilon_{I}, \epsilon_{Id}, \epsilon_{00}, \epsilon_{dd}, \epsilon_{II}$ vanish.
Then Eqs. (\ref{Ejl}) and (\ref{E22}) or (\ref{E221}) give the unexcited aether wave (transverse $\epsilon_{jl}\neq0$ and traceless metric $\epsilon_{II}=0$ ) modes---spin-2
\begin{eqnarray}\label{}
s^2_2=\frac{1}{1-c_{13}},
\end{eqnarray}
which is the same as that in 4-dimensional spacetime \cite{jacobson4}.
The constant $c_{13}$ should be less than unit to insure that $s_2^2>0$, here we employ the constrain
\begin{eqnarray}\label{}
0\leq c_{13}<1.
\end{eqnarray}
Note that it is consistent to gost-free condition (the coefficients of the time kinetic term of each excitation $q_{S,V,T}$ must be positive, where $S,V,T$ are to be scalar, vector and tensor, respectively) $q_T=1-c_{13}>0$ \cite{oost}.

The second two modes correspond  to transverse aether-metric when polarizations $\epsilon_{I}$ and $\epsilon_{Id}$ are nonzero. Then Eqs. (\ref{AI}) and (\ref{EId}) give the transverse aether wave modes---spin-1
\begin{eqnarray}\label{}
s^2_1=\frac{2c_1-c_1^2+c_3^2}{2c_{14}(1-c_{13})},
\end{eqnarray}
which is also the same as that in 4-dimensional spacetime \cite{jacobson4}.

The fifth mode corresponds to nonzero polarizations $\epsilon_{00}, \epsilon_{dd}, \epsilon_{II}$. To avoid over determining the speed, the difference equation must be identically satisfied, i.e., the sums in parenthesis $()$ of Eq. (\ref{E22}) or (\ref{E221}) must be zero. Then Eqs. (\ref{Ad}, \ref{EII}, \ref{Edd}) give the trace aether wave modes---spin-0
\begin{eqnarray}\label{spin0}
s^2_0=\frac{c_{123}[n-(n-1)c_{14}]}{c_{14}(1-c_{13})[n(1+c_2)+(n-1)c_{123}]}.
\end{eqnarray}
It is only this speed of spin-0 wave mode that depends  on dimension number $n$.
As one can see that, when $c_{14}\rightarrow0$, the speed of spin-0 $s_0\rightarrow\infty$, means that the spin-0 horizon coincides with the universal horizon of a black hole. While $c_{123}\rightarrow0$, $s_0\rightarrow0$.
Eq. (\ref{tmass}) shows that $c_{14}<n/(n-1)$, and in the small $c_i$ limit, $s_0^2\rightarrow c_{123}/c_{14}$ implies that $c_{123}/c_{14}>0$. The gost-free condition $q_V=c_{14}>0$ \cite{oost} gives that $c_{123}>0$. Therefore here we employ the constrain
\begin{eqnarray}\label{}
0\leq c_{14}<\frac{n}{n-1}.
\end{eqnarray}
To insure that $s_0^2>0$, here we employ another constrain
\begin{eqnarray}\label{}
1+c_2+\frac{n-1}{n}c_{123}>0.
\end{eqnarray}

\section{Some derivations}
In this section, we give derivations for some formulas in maintext.

{\bf A}. For the first line of Eq. (\ref{EMTs}), it can be obtained by varying the action with respect to metric $g_{ab}$,
\begin{eqnarray}\label{deltag}
-\frac{1}{\sqrt{-g}}\delta_g(\sqrt{-g}\mathcal{L}_{\ae})
=\frac{1}{2}g_{ab}\mathcal{L}_{\ae}\delta g^{ab}+(\delta_g Z^{ab}_{~~~cd})
(\nabla_au^c)(\nabla_bu^d)-\lambda\delta_gu^2+Y^a_{~~c}\delta_g(\nabla_au^c)
+Y^b_{~~d}\delta_g(\nabla_bu^d).
\end{eqnarray}
The first term on the right side of Eq. (\ref{deltag}) is
\begin{eqnarray}\label{}
\frac{1}{2}g_{ab}\mathcal{L}_{\ae}\delta g^{ab}=-\frac{1}{2}g_{ab}Y^c_{~~d}(\nabla_cu^d)\delta g^{ab},
\end{eqnarray}
where the constraint $u^2=-1$ is used.
By using
\begin{eqnarray}\label{}
\delta_g Z^{ab}_{~~~cd}=c_1(g_{cd}\delta g^{ab}+ g^{ab}\delta g_{cd})-c_4u^au^b\delta g_{cd},
\end{eqnarray}
we can obtain the second term on the right side of Eq. (\ref{deltag}) is
\begin{eqnarray}\label{}
(\delta_g Z^{ab}_{~~~cd})(\nabla_au^c)(\nabla_bu^d)
=c_1[(\nabla_au_c)(\nabla_bu^c)-(\nabla^cu_a)(\nabla_cu_b)]\delta g^{ab}+c_4a_aa_b\delta g^{ab}.
\end{eqnarray}
The third term on the right side of Eq. (\ref{deltag}) is
\begin{eqnarray}\label{}
-\lambda\delta_gu^2=-\lambda u^{\sigma}\delta_gu_\sigma=-\lambda u^{\sigma}u^\rho\delta g_{\sigma\rho}=\lambda u_au_b\delta g^{ab},
\end{eqnarray}
where $\delta_g u^\sigma=0$ is used. The fourth and the fifth term on the right side of Eq. (\ref{deltag}) are identical due to the symmetry $Z^{ba}_{~~~dc}=Z^{ab}_{~~~cd}$. By using the variation of the Christoffel \cite{carroll}
\begin{eqnarray}\label{}
\delta_g\Gamma_{a\tau}^c=-\frac{1}{2}\big(g_{ar}\nabla_\tau\delta g^{cr}+g_{\tau r}\nabla_a\delta g^{cr}+g^{cr}\nabla_r\delta g_{a\tau}\big),
\end{eqnarray}
we can obtain the fourth term is
\begin{eqnarray}\label{}
Y^a_{~~c}\delta_g(\nabla_au^c)=\frac{1}{2}\nabla_aX^a_{~cr}\delta g^{cr}=\frac{1}{2}\nabla_cX^c_{~ab}\delta g^{ab}.
\end{eqnarray}
Putting everything together, one can obtain the Eq. (\ref{EMTs}) for the stress-energy tensor $\mathcal{T}^{\ae}_{ab}$.

{\bf B}. For the eqs. (\ref{IDa}), some coefficients of the projections of $\mathcal{T}^{\ae}_{ab}$ and $\mathcal{T}^{M}_{ab}$ on the aether frame are needed,
\begin{eqnarray}
&&\mathcal{T}^{\ae}_{uu}=c_{14}(\nabla\cdot a-a^2)+\frac{1}{2}\bar L_{\ae},\;\;\bar L_{\ae}=c_{14}a^2-c_2K^2-c_{13}K_{cd}K^{cd},\nonumber\\
&&\mathcal{T}^{\ae}_{ss}=c_{13}(\nabla_uK_0+KK_0)-c_{14}a^2+c_2(K^2+\nabla_uK)+\frac{1}{2}\bar L_{\ae},\nonumber\\
&&\hat{\mathcal{T}}^{\ae}=c_{13}(\nabla_u\hat{K}+K\hat{K})+nc_2(K^2+\nabla_uK)+\frac{n}{2}\bar L_{\ae},\nonumber\\
&&\mathcal{T}^{\ae}=g^{ab}\mathcal{T}^{\ae}_{ab}=-\mathcal{T}^{\ae}_{uu}
+\mathcal{T}^{\ae}_{ss}+\hat{\mathcal{T}}^{\ae},\nonumber\\
&&\mathcal{T}^{M}_{ss}=-\mathcal{T}^{M}_{uu}=\frac{\tilde{Q}^2}{8\pi G_{\ae}r^{2n}},\;\hat{\mathcal{T}}^{M}=-\frac{n\tilde{Q}^2}{8\pi G_{\ae}r^{2n}},
\end{eqnarray}
which are the inner products of $\mathcal{T}^{\ae}_{ab}$ and $\mathcal{T}^{M}_{ab}$ with the $u^au^b,\;u^as^b,\;s^au^b,\;s^as^b,\;\hat g^{ab}$.
By using the definition of Riemann tensor $\mathcal{R}_{ab}u^b=\nabla^b(\nabla_au_b)-\nabla_a(\nabla\cdot u)$ and $\mathcal{R}_{ab}s^b=\nabla^b(\nabla_as_b)-\nabla_a(\nabla\cdot s)$, other coefficients of the projections of $\mathcal{R}_{ab}$ on the aether frame are
\begin{eqnarray}
&&\mathcal{R}_{uu}=\nabla\cdot a-\nabla_uK-(K_0^2+\frac{\hat K^2}{n}),\nonumber\\
&&\mathcal{R}_{ss}=\nabla_uK_0+KK_0-(a^2+\frac{\hat k^2}{n})-\nabla_s[(a\cdot s)+\hat k],
\end{eqnarray}
Then do the inner product of $\mathcal{T}^{\ae}_{ab}$ with the Killing vector $\chi^b$ as
\begin{eqnarray}
\label{}
\left(\mathcal{T}^{\ae}_{ab} - \frac{1}{n}g_{ab}\mathcal{T}^{\ae}\right)\chi^b &=&  \mathcal{T}^{\ae}_{uu}(u\cdot\chi)u_a
-\mathcal{T}^{\ae}_{us}[(s\cdot\chi)u_a+(u\cdot\chi)s_a]
+\mathcal{T}^{\ae}_{ss}(s\cdot\chi)s_a
-\frac{1}{n}\mathcal{T}^{\ae}\chi_a\nonumber\\
&=&\big(\mathcal{T}^{\ae}_{uu}-\mathcal{T}^{\ae}_{ss}-\frac{\bar L_{\ae}}{2}\big)(u\cdot\chi)u_a
+\big(\mathcal{T}^{\ae}_{ss}-\mathcal{T}^{\ae}_{uu}+\frac{\bar L_{\ae}}{2}\big)(s\cdot\chi)s_a-\frac{1}{n}\mathcal{T}^{\ae}\chi_a\nonumber\\
&=&\big(\mathcal{T}^{\ae}_{ss}-\mathcal{T}^{\ae}_{uu}+\frac{\bar L_{\ae}}{2}-\frac{1}{n}\mathcal{T}^{\ae}\big)\chi_a\nonumber\\
&=&\nabla_c\big[-(1-\frac{1}{n})c_{14}a^c+c_{13}K_0u^c-\frac{c_{123}}{n}Ku^c\big]\chi_a.
\end{eqnarray}
By using $(\nabla\cdot V)\chi_a=\nabla^b(V_\rho u_{[a}s_{b]})$ \cite{bhattacharyya2}, where $V_\rho=V\cdot\rho$ and $(\rho\cdot\chi)=0,\;\rho^2=-\chi^2$, one can get
\begin{eqnarray}
\label{}
\left(\mathcal{T}^{\ae}_{ab} - \frac{1}{n}g_{ab}\mathcal{T}^{\ae}\right)\chi^b = \nabla^b\Big\{\big[\big(\frac{c_{123}}{n} K  - c_{13} K_0\big) (s\cdot \chi)+(1-\frac{1}{n})c_{14}(a\cdot s)(u\cdot\chi)\big]u_{[a}s_{b]}\Big\} .
\end{eqnarray}
Similarly, do the inner product of $\mathcal{T}^{M}_{ab}$ with the Killing vector $\chi^b$ as
\begin{eqnarray}
\label{}
8\pi G_{\ae}\left(\mathcal{T}^{M}_{ab} - \frac{1}{n}g_{ab}\mathcal{T}^{M}\right)\chi^b &=&8\pi G_{\ae}\Big\{  \mathcal{T}^M_{uu}(u\cdot\chi)u_a
-\mathcal{T}^M_{us}[(s\cdot\chi)u_a+(u\cdot\chi)s_a]
+\mathcal{T}^M_{ss}(s\cdot\chi)s_a
-\frac{1}{n}\mathcal{T}^M\chi_a\Big\}\nonumber\\
&=&\frac{2(n-1)\tilde{Q}^2}{nr^{2n}}\chi_a.
\end{eqnarray}
By using $2(n-1)\tilde{Q}^2/nr^{2n}=(\nabla\cdot V)=(r^nV_\rho)'/r^n$, where $'\equiv d/dr$, one can obtain
\begin{eqnarray}
\label{}
V_\rho=\frac{2\tilde{Q}^2}{nr^{n-1}}.
\end{eqnarray}
By using $(\nabla\cdot V)\chi_a=\nabla^b(V_\rho u_{[a}s_{b]})$,  one can get
\begin{eqnarray}
\label{}
8\pi G_{\ae}\left(\mathcal{T}^{M}_{ab} - \frac{1}{n}g_{ab}\mathcal{T}^{M}\right)\chi^b = \nabla^b\Big(\frac{2\tilde{Q}^2}{nr^{n-1}}u_{[a}s_{b]}\Big) .
\end{eqnarray}

{\bf C}. For the solution (\ref{aether}), the $uu-$ component of the gravitational field equation (\ref{uussA}) becomes
\begin{eqnarray}
\label{}
\nabla\cdot a-(1-c_{13})K_{ab}K^{ab}+2(\frac{1}{n}-1)\frac{\tilde{Q}^2}{r^{2n}}=0,
\end{eqnarray}
where $\nabla\cdot a=\nabla_s(a\cdot s)+(a\cdot s)^2+(a\cdot s)\hat k,\;K_{ab}K^{ab}=K_0^2+\hat K^2/n$. By using Eqs. (\ref{AS}) and $(s\cdot\chi)=r^n_{\ae}/r^n$, one can obtain that
\begin{eqnarray}
\label{}
\frac{1}{r^n}[r^n(u\cdot\chi)(u\cdot\chi)']'
=(1-c_{13})n(n+1)\frac{r_{\ae}^{2n}}{r^{2n+2}}
-2(\frac{1}{n}-1)\frac{\tilde{Q}^2}{r^{2n}}.
\end{eqnarray}
One can solve it and get
\begin{eqnarray}
\label{}
(u\cdot\chi)^2=(1-c_{13})\frac{r_{\ae}^{2n}}{r^{2n}}
+\frac{2}{n(n-1)}\frac{\tilde{Q}^2}{r^{2n-2}}
-\frac{t_1}{(n-1)r^{n-1}}+t_2.
\end{eqnarray}
Since $(u\cdot\chi)\rightarrow-1$ at infinite, one can set $t_1/(n-1)=\bar r_0,\;t_2=1$, $(u\cdot\chi)=-\sqrt{(u\cdot\chi)^2}$ and can obtain Eq. (\ref{aether}).

{\bf D}. For the solution (\ref{metric2}), the $uu-$ component of the gravitational field equation (\ref{uussA}) becomes
\begin{eqnarray}
\label{}
\big[1-(1-\frac{1}{n}c_{14})\big]\nabla\cdot a-(1-c_{13})(\nabla _uK+K_{ab}K^{ab})+2(\frac{1}{n}-1)\frac{\tilde{Q}^2}{r^{2n}}=0.
\end{eqnarray}
Also by using Eqs. (\ref{AS}) and $(u\cdot\chi)=-1+\bar r_0/2r^{n-1}$, one can obtain that
\begin{eqnarray}
\label{}
(1-c_{13})\frac{1}{r^n}[r^n(s\cdot\chi)(s\cdot\chi)']'
=\big[1-(1-\frac{1}{n}c_{14})\big]\frac{(n-1)^2\bar r_0^2}{4r^{2n}}
+2(\frac{1}{n}-1)\frac{\tilde{Q}^2}{r^{2n}}.
\end{eqnarray}
One can solve it and get
\begin{eqnarray}
\label{}
(1-c_{13})(s\cdot\chi)^2=\big[1-(1-\frac{1}{n}c_{14})\big]\frac{\bar r_0^2}{4r^{2n-2}}
-\frac{2}{n(n-1)}\frac{\tilde{Q}^2}{r^{2n-2}}
-\frac{t_1}{(n-1)r^{n-1}}+t_2.
\end{eqnarray}
Since $(s\cdot\chi)\rightarrow0$ at infinite, one can set $t_2=0$. Another constant $t_1$ is also zero, because $(s\cdot\chi)$ should be analytic in $1/r^{n-1}$ at infinity and captured by the standard boundary condition imposed on the aether. Lastly one can obtain Eq. (\ref{metric2}).

\end{document}